\documentclass[a4paper,12pt]{article}
\usepackage{cite}
\usepackage{soul}
\usepackage{amssymb}
\usepackage{multicol}
\usepackage{amsmath}
\usepackage{pgfplots}
\usepackage{mathrsfs} 
\usepackage{xfp} 
\usepackage[outline]{contour} 
\usepackage[d]{esvect}

\usepackage{epsfig}
\usepackage{graphicx}
\usepackage{slashed}
\usepackage{xcolor,color}
\usepackage{ulem}
\usepackage{subfigure}
\usepackage[toc,page]{appendix}
\usepackage[makeroom]{cancel}
\usepackage{tikz}
\usetikzlibrary{shapes.geometric,positioning,decorations.pathreplacing,decorations.pathmorphing,patterns,decorations.markings}

\usepackage{scalerel}
\usepackage[usestackEOL]{stackengine}

\def\dashint{\,\ThisStyle{\ensurestackMath{%
            \stackinset{c}{.2\LMpt}{c}{.5\LMpt}{\SavedStyle-}{\SavedStyle\phantom{\int}}}%
        \setbox0=\hbox{$\SavedStyle\int\,$}\kern-\wd0}\int}

\makeatletter

\@addtoreset{equation}{section}
\makeatother
\def\be{\begin{equation}}
\def\ee{\end{equation}}
\def\bea{\begin{eqnarray}}
\def\eea{\end{eqnarray}}
\def\bgn{\begin{align}}
\def\egn{\end{align}}

\def\({\left(}
\def\){\right)}
\def\<{\left<}
\def\>{\right>}

\def\({\left(}
\def\){\right)}
\def\<{\left<}
\def\>{\right>}
\def\!{\right|}
\def\|{\left|}

\def\[{\left[}
\def\]{\right]}

\def\+{\bar}

\def\N{{\cal{N}}}

\def\C{{\cal{C}}}

\def\K{{\cal{K}}}

\usepackage{hyperref}
\hypersetup{colorlinks,%
  citecolor=blue,%
  linkcolor=blue,%
  pdftex}

\definecolor{Sdark}{RGB}{66,76,92}
\definecolor{SmidDark}{RGB}{104,117,136}
\definecolor{Smid}{RGB}{153,166,184}
\definecolor{Slight}{RGB}{211,220,232}
\definecolor{Shigh}{RGB}{246,249,252}
\definecolor{Sedge}{RGB}{34,54,84}

\pgfdeclareverticalshading{surfacebody3d}{180bp}{
  color(0bp)=(Sdark);
  color(14bp)=(Sdark!78!SmidDark);
  color(30bp)=(SmidDark);
  color(48bp)=(SmidDark!58!Smid);
  color(67bp)=(Smid);
  color(86bp)=(Smid!48!Slight);
  color(105bp)=(Slight);
  color(123bp)=(Slight!38!Shigh);
  color(139bp)=(Shigh);
  color(154bp)=(Slight!62!Shigh);
  color(168bp)=(Slight!58!Smid);
  color(180bp)=(Smid)}

\pgfdeclareradialshading{softlight}{\pgfpoint{0cm}{0cm}}{
  color(0bp)=(white);
  color(18bp)=(white!96!Shigh);
  color(38bp)=(Shigh);
  color(62bp)=(Slight!55!Shigh);
  color(82bp)=(Slight);
  color(100bp)=(Smid!74!Slight)}
\pgfdeclareradialshading{softshadow}{\pgfpoint{0cm}{0cm}}{
  color(0bp)=(Sdark);
  color(20bp)=(Sdark!72!SmidDark);
  color(45bp)=(SmidDark);
  color(72bp)=(SmidDark!58!Smid);
  color(100bp)=(Smid)}

\tikzset{pics/puncture3d/.style={code={%
    \path[fill=white,draw=none]
      (-.335,0)
      .. controls (-.180,.103) and (.180,.103) .. (.335,0)
      .. controls (.175,-.082) and (-.175,-.082) .. (-.335,0) -- cycle;
    \draw[Sedge!90!black,line width=.31pt,line cap=round]
      (-.335,0)
      .. controls (-.175,.105) and (.175,.105) .. (.335,0);
    \draw[Sedge!90!black,line width=.31pt,line cap=round]
      (-.335,0)
      .. controls (-.178,-.087) and (.178,-.087) .. (.335,0);
    \draw[Sedge!90!black,line width=.31pt,line cap=round]
      (-.335,0) .. controls (-.361,.021) and (-.371,.055) .. (-.379,.084);
    \draw[Sedge!90!black,line width=.38pt,line cap=round]
      (.335,0) .. controls (.357,.017) and (.362,.041) .. (.366,.059);
}}}

\tikzset{pics/puncture3dcompact/.style={code={%
    \path[fill=white,draw=none]
      (-.335,0)
      .. controls (-.180,.103) and (.180,.103) .. (.335,0)
      .. controls (.175,-.082) and (-.175,-.082) .. (-.335,0) -- cycle;
    \draw[Sedge!80!black!50,line width=.31pt,line cap=round]
      (-.335,0)
      .. controls (-.175,.105) and (.175,.105) .. (.335,0);
    \draw[Sedge!90!black,line width=.31pt,line cap=round]
      (-.335,0)
      .. controls (-.178,-.087) and (.178,-.087) .. (.335,0);
    \draw[Sedge!80!black,line width=.31pt,line cap=round]
      (-.335,0) .. controls (-.361,.021) and (-.371,.055) .. (-.379,.085);
    \draw[Sedge!90!black,line width=.3pt,line cap=round]
      (.335,0) .. controls (.357,.017) and (.362,.041) .. (.366,.059);
}}}

\begin{document}

\begin{titlepage}
\vskip1cm
\begin{flushright}
\end{flushright}
\vskip0.25cm
\centerline{
\bf \large 
Confining Random Potential Quantum Mechanics for Jackiw–Teitelboim Gravity
} 
\vskip0.8cm \centerline{ \textsc{
 Dongsu Bak,$^{ \negthinspace  a}$  Chanju Kim,$^{ \negthinspace b}$ Sang-Heon Yi,$^{\negthinspace a}$} }
\vspace{0.8cm} 
\centerline{\sl  a) Physics Department \& Natural Science Research Institute}
\centerline{\sl University of Seoul, Seoul 02504 \rm KOREA}
 \vskip0.2cm
 \centerline{\sl b) Department of Physics, Ewha Womans University,
  Seoul 03760 \rm KOREA}
\vskip0.4cm

 \centerline{
\tt{(\small dsbak@uos.ac.kr,\,cjkim@ewha.ac.kr,\,shyi704@uos.ac.kr})
} 
  \vspace{1.5cm}
\centerline{ABSTRACT} \vspace{0.65cm}
{
\noindent
We formulate perturbative Jackiw-Teitelboim gravity as an ensemble of ordinary
one-dimensional Hamiltonians with a confining random potential.  
The leading profile reproduces the disk density of states, while connected cumulants 
of a single master field encode the topological expansion.  After formulating the 
cumulant power counting required by the topological expansion, we show that the cumulants reproduce all genus-zero multi-boundary amplitudes $\mathcal Z_{0,n}$.
At higher genus, we show that a renormalization scheme can be chosen 
consistently order by order so that imposing one-boundary polynomiality 
recursively reproduces the perturbative JT string equation and yields 
the standard JT one-boundary amplitudes $\mathcal Z_{g,1}$. The framework provides a self-contained   finite-$S_0$ 
quantum dynamics with a discrete spectrum and unitary real time evolution,
beyond a mere rewriting of the matrix integral.  The same confining 
dynamics also underlies late-time saturation and the dynamical avoidance 
of the would-be black hole singularity, as shown in our earlier work.
}

\end{titlepage}

\section{Introduction}\label{sec:introduction}

Jackiw-Teitelboim (JT) gravity~\cite{Jackiw:1984je,Teitelboim:1983ux,Almheiri:2014cka}
admits a Euclidean topological expansion of the connected multi-boundary
correlators of partition function operators. This expansion is reproduced by
 the Saad-Shenker-Stanford (SSS) matrix theory~\cite{Saad:2019lba}.
On the Lorentzian side, the leading Schwarzian theory is an ordinary quantum mechanics \cite{Jensen:2016pah,Maldacena:2016upp, Engelsoy:2016xyb,Bagrets:2016cdf,  Mertens:2017mtv,Stanford:2017thb,Kitaev:2017awl, Harlow:2018tqv, Penington:2023dql}, in which the negative of the coordinate $q$ is interpreted as the renormalized geodesic length connecting the left and right boundaries. However, the Schwarzian theory has a continuous spectrum, which is in tension with 
the expectation that a finite-entropy black hole should have a discrete spectrum~\cite{Maldacena:2001kr}.

To resolve this tension, we add a confining random potential $W(X)$ to the Schwarzian system \cite{Bak:2025eul}, where $X=-q e^{-S_0}$ and $e^{-S_0}=\hbar$ sets the average level-spacing scale. The resulting Hamiltonian is  $H=p_X^2+e^{-X/\hbar}+W(X)$.  The additional confining potential $W(X)$ becomes $O(1)$ only for $q=-O(e^{S_0})$  
and therefore decouples in the  $S_0\rightarrow\infty$ limit. We decompose it as $W(X)=W_0(X)+\hbar w(X)$,  where the deterministic part $W_0$ is completely fixed by the disk density of states \cite{Bak:2025eul}, while the single master field
$w(X)=\sum_{a=1}^{\infty}\hbar^{a-1}w_a(X)$
encodes the  multi-boundary and higher-genus
cumulant data. In Ref.~\cite{Bak:2026tvy}, we showed that 
this additional confining dynamics  resolves the black-hole singularities and renders the bulk geometry horizonless.

The main advantage of our Hamiltonian formulation lies in studying the real time JT dynamics. Once a  potential  is drawn from the ensemble,
 it defines
an ordinary one-dimensional Hamiltonian with its own Hilbert space, discrete spectrum, and unitary time evolution of the corresponding JT system. For example, the geodesic length $\ell=-q$ can be identified with the Krylov spread complexity of states~\cite{Susskind:2018pmk,Susskind:2014rva,Susskind:2014moa,Brown:2018bms,Susskind:2019ddc,Susskind:2020gnl,Iliesiu:2021ari,Lin:2022rbf,Balasubramanian:2022tpr,Rabinovici:2023yex,Bak:2025qgs}, and the resulting time evolution plays a crucial role in the dynamical resolution of the black-hole singularity \cite{Bak:2026tvy}. Our Hamiltonian formulation is therefore particularly suited to questions for which fixed-realization dynamics is important, rather than merely serving as an alternative way of organizing ensemble averaged Euclidean correlators. For real time observables, the natural prescription is correspondingly quenched: one first fixes a realization, evolves it unitarily, and only afterwards performs the ensemble average.

In this paper, we test the perturbative Hamiltonian formulation in two complementary sectors of the JT topological expansion: arbitrary boundary number at genus zero and arbitrary genus with one boundary. We
 show that the resulting correlators agree with those obtained from the topological expansion of JT gravity (see~\cite{Mertens:2022irh,Turiaci:2024cad} for a review). First, we note that the partition function operator admits the standard Wigner-Kirkwood (WK) expansion in $\hbar$~\cite{Wigner:1932eb}. 
Requiring the connected cumulants of $w$ and its composites to 
respect the JT genus grading then fixes their allowed $\hbar$ scaling. 
Together with the WK expansion, this  is consistent with the gravitational weight 
$\hbar^{2g+n-2}$ and determines which genera can contribute to each cumulant channel.

Second, we shall work out the genus-zero sector completely. We start with a specific form for the genus-zero $n$-boundary primitive cumulants of the single master field and assume that all remaining genus-zero cumulants of composite operators are constructed from these primitive cumulants by a simple point-splitting rule. Within this setup, we show that the complete family of known amplitudes $Z_{0,n}$ can be reconstructed, which provides a nontrivial test of our formulation. Third, we shall analyze the one-boundary sector completely at arbitrary genus.
There we show that a renormalization scheme can be chosen consistently order by order so that one-boundary polynomiality in $\beta$ recursively reproduces the perturbative JT string equation and yields the standard JT one-boundary amplitudes $Z_{g,1}(\beta)$ with $g\ge 1$.

We may proceed  to the remaining sector, $Z_{g,n}$ with $g>0$ and $n>1$. There is, however, a renormalization-scheme dependence in the local description of the relevant cumulants. Individual coincident composite operators are not unique: operator mixing and finite local counterterms can redistribute local contact terms without changing the desired partition function correlators. Rather than constructing the higher-genus cumulants directly within this scheme-dependent local description, we therefore plan to pursue a different route, which will be the subject of future work.

The paper is organized as follows.  Section~\ref{sec:boundary-dynamics} reviews the leading JT  Schwarzian dynamics. Section~\ref{sec:ensemble-qm} introduces the ensemble of confining Hamiltonians and the single master field description.  Section~\ref{sec:topological-expansion} derives the topological power counting of the cumulants.  Section~\ref{sec:Z0n} reconstructs all genus-zero multi-boundary amplitudes, Section~\ref{sec:one-boundary} analyzes the one-boundary sector at arbitrary genus, and Section~\ref{sec:conclusions} summarizes the results and open problems.  Appendix~\ref{app:WK} derives the WK expansion used in the
 main text.

\section{JT gravity and boundary quantum mechanics}\label{sec:boundary-dynamics}

We first fix our conventions for JT gravity and summarize how the one-dimensional quantum mechanics arises.
The bulk gravity action with the Lorentzian signature is
\begin{equation}
I=\frac{1}{16\pi G}\int_M d^2x\sqrt{-g}\,\phi(R+2)
 +\frac{1}{8\pi G}\int_{\partial M}dt\sqrt{-\gamma_{tt}}\,\phi(K-1)\,,
\end{equation}
where the boundary term denotes both asymptotic boundaries in the two-sided geometry.  Its bulk equations of motion are
\begin{equation}
R+2=0\,,\qquad
\nabla_a\nabla_b\phi-g_{ab}\nabla^2\phi+g_{ab}\phi=0\,.
\end{equation}
The first equation fixes the metric locally to AdS$_2$.  The Euclidean theory also contains the topological term
\begin{equation}
I_E^{\rm top}=-\frac{S_0}{2\pi} 
\left[\frac12\int_{M_E}\sqrt g\,R+\int_{\partial M_E}\sqrt\gamma\,K\right]\,,
\end{equation}
which weights a connected surface of genus $g$ with $n$ asymptotic boundaries by
$e^{-(2g-2+n)S_0}$.  Here and below, we set $8\pi G=1$ just for convenience. 

The Euclidean path integral sums over connected surfaces of different topology, and the powers of $e^{-S_0}$ organize the perturbative expansion by the Euler characteristic.  In the SSS 
description~\cite{Saad:2019lba}, the amplitudes $Z_{g,n}$ are reproduced by connected correlators of matrix partition function operators. 
This matrix formulation efficiently determines the perturbative spectral observables, while the Lorentzian boundary dynamics provides the direct geometric interpretation of 
phase-space variables 
in the reduced Hamiltonian.

To obtain the Hamiltonian, let  us establish  our conventions.
In global coordinates,  the metric of AdS$_{2}$ is  given by
\begin{equation}\label{global}
ds^2=\frac{-d\tau^2+d\mu^2}{\cos^2\mu}\,,\qquad -\frac\pi2\leq\mu\leq\frac\pi2\,.
\end{equation}
The cutoff trajectory curve is specified by
\begin{equation}\label{cutoff}
\phi\big|_c=\frac{\C 
}{\epsilon},\qquad
 ds^2\big|_c=-\frac{dt^2}{\epsilon^2}\,,
\end{equation}
where $\epsilon \ll 1$ is a cutoff parameter.
Then,
the leading boundary behavior of the dilaton directly determines the physical time reparametrization.
The conventional holographic computation gives us the Schwarzian action on either boundary:
\begin{equation}
L_{r/l}=-\mathcal C\,\mathrm{Sch}\left(\tan\frac{\tau_{r/l}}2,t\right)\,,
\end{equation}
with $\mathrm{Sch}(f,t)=(f''/f')'-\frac12(f''/f')^2$ and $r/l$ denotes the right/left boundary, respectively.  Introducing auxiliary variables $\chi_{r/l}$ as  $e^\chi=\mathcal C\dot\tau$
 gives the first-order form
\begin{equation}
L_{r/l}=\frac{\mathcal C}{2}\dot\chi_{r/l}^{\,2}
-\frac{e^{2\chi_{r/l}}}{2\mathcal C}
+p_{r/l}\left(\dot\tau_{r/l}-\frac{e^{\chi_{r/l}}}{\mathcal C}\right).
\end{equation}
After imposing the two-sided $SL(2, {\bf R})$ constraints and fixing the large gauge redundancy, the whole Lagrangian reduces to
\begin{equation}\label{total}
L_{\rm total}=\mathcal C\dot\chi^{\,2}-\frac{e^{2\chi}}{\mathcal C}\,.
\end{equation}
Let $p_\chi=2\mathcal C\dot\chi$ be the momentum conjugate to $\chi$ in
$L_{\rm total}$.  
Introducing also $q=2\chi$ with its conjugate momentum
$p_q=p_\chi/2$, one obtains
\begin{equation}\label{SchHam}
2\mathcal C H=\frac14 p_\chi^2+e^{2\chi}=p_q^2+e^q\,,
\qquad q=2\chi\,, \qquad p_q=\frac12 p_\chi\,.
\end{equation}
Here $H=H_{\mathrm{total}}/2$ is the Hamiltonian of either boundary.
Note that  $\chi$ variable can be identified with  the renormalized two-sided length as $\ell_{\rm ren}=-2\chi =-q$~\cite{Harlow:2018tqv}.  The Liouville potential $e^q$ 
grows without bound only as  $q\to+\infty$; as $q\to-\infty$ it vanishes, so the Schwarzian Hamiltonian describes a physical system composed of   scattering states with a continuous spectrum.

On the other hand, the JT disk density of states is~\cite{Stanford:2017thb}
\begin{equation}
\rho_{\rm JT}(E)=\frac{e^{S_0}}{4\pi^2}\sinh(2\pi\sqrt E)\,,
\end{equation}
which is smooth and finite at leading order, whereas the unbounded $q\to-\infty$ region of the Schwarzian system gives rise to an infinite density of scattering states.
This motivates us to complete the Schwarzian quantum mechanics by a left confining potential acting at $q=-O(e^{S_0})$.  The Liouville wall then controls the usual semiclassical JT region, while the additional, parametrically remote wall becomes relevant only at exponentially large wormhole length, where spectral discreteness and finite-entropy effects can no longer be ignored.   Our aim in later sections is to combine these two ingredients in a local one-dimensional Hamiltonian whose potential fluctuates randomly across realizations.
\section{Quantum-mechanical formulation with a confining random potential}\label{sec:ensemble-qm}

The SSS matrix model and Schwarzian descriptions of JT gravity emphasize complementary aspects of the same physics. The matrix model determines the perturbative connected multi-boundary amplitudes, but its eigenvalue variables do not have a direct geometric interpretation. 
The Schwarzian theory, by contrast, makes the Lorentzian boundary dynamics manifest, but its Liouville potential is nonconfining in one direction, resulting in a continuous spectrum, and by itself carries no information about the ensemble correlations.
In this section we review our formulation~\cite{Bak:2025eul} that combines the
two approaches: an ensemble of local one-dimensional Hamiltonians with a confining random
potential.  Here, each realization is an ordinary quantum system, the coordinate
retains its JT interpretation through $X=-\hbar q$, and the connected correlation functions of a local random variable have  the topological expansion. To avoid confusion with the matrix model correlators, these connected random variable correlation functions will be
called cumulants in the following.

Let $M$ be the double-scaled Hermitian matrix of the JT ensemble, with
$\widehat Z(\beta)=\mathrm{tr}\,e^{-\beta M}$.  Its connected correlators have
the perturbative expansion~\cite{Saad:2019lba,Mehta2004,Eynard:2015aea}
\begin{equation}\label{genusn}
\big\langle\widehat Z(\beta_1)\cdots\widehat Z(\beta_n)\big\rangle_{\rm conn}
\cong\sum_{g=0}^\infty \hbar^{2g+n-2}Z_{g,n}(\beta_1,\ldots,\beta_n)\,,
\end{equation}
where $\cong$ denotes equality up to terms nonperturbative in $\hbar$.  The
goal of the quantum mechanical formulation is to reproduce this expansion by
averaging over a random confining potential.

On the quantum mechanical side we use the rescaled coordinate
\begin{equation}
X=-\hbar\, q\,,
\label{Xrescaling}
\end{equation}
and, setting $2\mathcal C=1$, the Hamiltonian of Section~\ref{sec:boundary-dynamics}
becomes
\begin{equation}\label{QMHamiltonian}
H=p_X^2+V(X)\,,\qquad p_X=-i\hbar\partial_X\,,
\qquad V(X)=e^{-X/\hbar}+W(X)\,,
\end{equation}
where the first term in $V(X)$ is the original Liouville wall and $W(X)$ is
the random confining potential that incorporates the matrix-model randomness
into the Hamiltonian.  
Following Ref.~\cite{Bak:2026tvy}, we may decompose $W(X)$ into a deterministic profile $W_0(X)$ and a fluctuating part $\hbar w(X)$ as 
\begin{equation}\label{masterfield}
W(X)=W_0(X)+\sum_{m=1}^\infty\hbar^m w_m(X)
      \equiv W_0(X)+\hbar w(X)\,.
\end{equation}
We shall mainly use the single master field $w \, (=\sum_{m=1}^\infty\hbar^{m-1}w_m(X))$, which keeps the topological bookkeeping compact, while the component fields $w_m$ remain useful for isolating individual orders in $\hbar$.  For the partition functions,  it is convenient to introduce
\begin{equation}
Z(\beta)\equiv \mathrm{tr}\,e^{-\beta H}\,,\qquad
\mathcal Z(\beta)=\left(\frac{\pi}{\beta}\right)^{1/2}Z(\beta)\,,
\label{traceDefinitions}
\end{equation}
and correspondingly, for the amplitudes appearing in \eqref{genusn},
\begin{equation}
\mathcal Z_{g,n}=\left(\frac{\pi^n}{\beta_1\cdots\beta_n}\right)^{1/2}Z_{g,n}\,.
\label{amplitudeRescaling}
\end{equation}

The introduction of a confining potential $W_{0}$  has a direct physical motivation.  As emphasized in
Ref.~\cite{Bak:2026tvy}, the pure Schwarzian Hamiltonian has a continuous
spectrum because the Liouville potential is open toward $q\to-\infty$,
whereas a finite-entropy black hole should have a discrete spectrum with mean
level spacing of order $e^{-S_0}$.  The left confining potential wall closes off the open
direction and converts the scattering problem into a bound-state problem at
 finite $S_0$.  Its effect is negligible in the ordinary
semiclassical region and becomes important only at $q=-O(e^{S_0})$,
equivalently when the renormalized wormhole length $\ell_{\rm ren}=-q$
reaches $O(e^{S_0})$.  Since $-q$ can be identified with the Krylov spread
complexity in this setting~\cite{Susskind:2018pmk,Susskind:2014rva,Susskind:2014moa,Brown:2018bms,Susskind:2019ddc,Susskind:2020gnl,Iliesiu:2021ari,Lin:2022rbf,Balasubramanian:2022tpr,Rabinovici:2023yex,Bak:2025qgs}, the same wall creates a turning region at
the scale $e^{S_0}$ and thereby accounts for the ramp-top-slope-plateau
behavior of the complexity. 
This turnaround dynamically resolves the bulk black hole singularity and renders the resulting bulk geometry horizonless~\cite{Bak:2026tvy}.

The deterministic profile $W_0$ follows from the disk amplitude.  For the
monotone half-line Hamiltonian $p_X^2+W_0(X)$, the reduced semiclassical
density is
\begin{align}
\frac{1}{4\pi^2}\sinh(2\pi\sqrt E)
 =\frac{1}{2\pi}\int_0^{X(E)}\frac{dX}{\sqrt{E-W_0(X)}} =\frac{1}{2\pi}\int_0^E\frac{du}{\sqrt{E-u}}\frac{dX}{du}
 \,.
 \label{AbelDensity}
\end{align}
Abel inversion of \eqref{AbelDensity} gives
\begin{equation}
 u=W_0(X)\,,\qquad
2\pi X=\sqrt u\,I_1(2\pi\sqrt u)\,.
\label{LCPot}
\end{equation}
Indeed, $dX/du=\tfrac12 I_0(2\pi\sqrt u)$, which,  when inserted in
\eqref{AbelDensity}, reproduces the stated JT density  of states.  The map is monotonic
for $X,u\geq0$ and satisfies $W_0(0)=0$. 
As discussed  in Appendix \ref{AppA}, 
we define the perturbative sector by integrating the bulk WK density over $X\geq0$, retaining $X=0$ as the endpoint. Separate boundary layer corrections from the Liouville wall $e^{-X/\hbar}$ are not included in this prescription.

We now proceed as follows. 
The formal disorder cumulants of the quantum mechanical ensemble are constrained by the matching condition
\begin{equation}\label{ensemblemap}
\big\langle\widehat Z(\beta_1)\cdots\widehat Z(\beta_n)\big\rangle 
\cong
\big\langle Z(\beta_1)\cdots Z(\beta_n)\big\rangle_{w}\,,
\end{equation}
where the left-hand side denotes the matrix-ensemble average, while the right-hand side denotes the disorder average over the random potential $w$.  
In this matching condition and the perturbative calculations below, $Z$ denotes the projected bulk trace of Section~\ref{sec:topological-expansion}, with separate wall corrections to the full thermal trace in~\eqref{traceDefinitions} excluded.
Some comments are in order.  First, the
condition is imposed order by order in $\hbar$: no exact nonperturbative
probability measure is assumed, and one may equivalently match the connected amplitudes $Z_{g,n}$ and reconstruct the full moments by the moment cumulant relations.
In this sense,  $W_0$ reproduces the
disk density of states, while the connected cumulants 
of $w$ carry all
higher-boundary and higher-genus data.  Second, the matching is not a
Gaussian approximation: the two-point cumulant fixes only the first
nontrivial connected amplitude, and every higher connected cumulant is
constrained independently, so the construction is an order-by-order inverse
problem for local cumulants rather than an assumed all-order recursion.
Third, individual coincident-point composite operator cumulants are not unique:
operator mixing, and  finite local counterterms 
can redistribute local
contacts. In our approach, the invariant objects to be matched are  the 
correlators in \eqref{ensemblemap}; there is freedom in choosing local
representatives as in the usual composite operator renormalization discussed in the introduction.

It is worth emphasizing where the randomness resides, because this is the structural difference between our present formulation and the matrix model (see the matrix model side, for example~\cite{Cotler:2016fpe, Saad:2018bqo,   Saad:2019lba, Maxfield:2020ale, Jafferis:2022wez}). In the matrix description the random variables are the eigenvalues themselves and hence the randomness is spectral. In our formulation, instead, the random variable is a  potential $W(X)$, while the spectrum of each realization is a nonlinear functional of $W$.  This shift of the randomness from the spectrum to the potential has two important consequences.  On the one hand, it is exactly what makes the formulation suitable for  dynamical problems: once a potential is drawn, one has a definite Hamiltonian with its own discrete levels and unitary evolution. 
On the other hand, the cumulants used here are represented by local contacts in $X$, rather than by correlations of eigenvalues.
 Consequently, the topological expansion is not inherited directly from eigenvalue combinatorics but must be established independently,  as we shall demonstrate in later sections. 
By construction, the two ensembles should agree on the perturbative partition function observables in \eqref{ensemblemap}.

Finally, products of the master field at coincident points are treated as
renormalized composite operators,
\begin{equation}
O_m(X):=[w^m(X)]_R\,,\qquad m\geq1\,,\qquad O_1=w\,,
\label{compositeDefinition}
\end{equation}
where $R$ denotes the common coincident-point renormalization prescription used in our calculation. 
Unless stated otherwise, all correlators of $w$ and $O_m$ below denote connected cumulants. Their $\hbar$ scaling is the central input of the topological expansion developed in the next section.

\section{Topological expansion}\label{sec:topological-expansion}

The central question in this section is how the topological genus grading of JT gravity is encoded in an $\hbar$ grading of the connected random potential cumulants.
 In the JT description, the power of $\hbar$ is fixed geometrically by the Euler characteristic of the connected surface.  In our 
quantum mechanical description, no topology is visible in a single realization of the potential.  The same information must therefore reappear in the $\hbar$ dependence of the connected cumulants of the random field.  This observation turns the genus expansion into a 
power-counting problem
of the cumulants. 

Basically two 
ingredients define the perturbative sector.
First, the 
WK expansion of a one-dimensional Hamiltonian contains only even powers of $\hbar$ as described in Appendix \ref{AppA}. Second, the Liouville term $e^{-X/\hbar}$ is nonperturbative for $X>0$ and becomes an infinitely steep wall for $X<0$ as $\hbar\to0$.  We define the perturbative genus sector by the steep-wall, projected bulk-trace prescription: the wall is replaced by the endpoint condition $X\geq0$ (see also Appendix \ref{AppA}), and one works with
\begin{equation}
 H_{\rm pert}=p_X^2+W(X)\,,\qquad X\in[0,\infty)\,.
\label{Hpert}
\end{equation}
Equation~(\ref{Hpert}) is only a shorthand for the projected bulk-trace prescription and is not meant to define an independent self-adjoint  Hamiltonian on a half-line.
The prescription here is not an independently imposed Dirichlet or Neumann boundary condition 
on a half-line.  Standard boundary Seeley-DeWitt coefficients~\cite{Vassilevich:2003xt}, or finite width boundary-layer data of the soft exponential wall, would constitute additional completion data and are not included in the bulk WK 
series used here.
For a fixed realization of $W$, the projected bulk trace  has the formal WK expansion 
\begin{equation}
 Z(\beta)\cong \frac{1}{\hbar\sqrt{4\pi\beta}}
 \int_0^\infty dX\,e^{-\beta W(X)}
 \left[1+\sum_{g\geq1}\hbar^{2g}A_{2g}[W,\beta](X)\right],
\label{WKmain}
\end{equation}
where $A_{2g}$ is a polynomial in derivatives of $W$. 
As mentioned already, the derivation of  the WK 
expansion and the role of the endpoint are reviewed in Appendix~\ref{app:WK}.  Equation~\eqref{WKmain} implies that 
the WK 
corrections themselves can only raise the order by $\hbar^{2g}$.  Consequently, once the lowest allowed power of a cumulant is fixed, all higher contributions generated by the quantum mechanics should fall 
into successive genera to be consistent.

Using $W=W_0+\hbar w$, the expansion of $e^{-\beta\hbar w}$ generates the renormalized composite insertions $O_m=[w^m]_R$ accompanied by $\hbar^m$.  Consider a connected cumulant containing $n$ such insertions,
 $\left\langle O_{m_1}(X_1)\cdots O_{m_n}(X_n)\right\rangle_{\rm conn}$.
The $n$ partition functions supply the universal phase-space factor $\hbar^{-n}$, while the insertions supply $\hbar^{s(\vec m)}$.  A connected JT surface of genus $g$ with $n$ boundaries carries the weight $\hbar^{2g+n-2}$.  Hence the cumulant must contribute precisely the missing power.  
Requiring the cumulant hierarchy to respect the JT genus grading term by term then gives
\begin{equation}
\left\langle O_{m_1}(X_1)\cdots O_{m_n}(X_n)\right\rangle_{\rm conn}
 =\sum_{g=g_{\min}(\vec m)}^\infty
 \hbar^{\alpha(g,\vec m)}
 C^{[g]}_{\vec m}(X_1,\ldots,X_n),
\label{cumulantgenus}
\end{equation}
where $\alpha(g,\vec m)=2g+2n-2-s(\vec m)$ and $C^{[g]}_{\vec m}=O(\hbar^0)$ with $s(\vec m)=\sum_{i=1}^n m_i$.  
 
Indeed, one has then
 $\hbar^{-n}\,\hbar^{s(\vec m)}\,
 \hbar^{\alpha(g,\vec m)}
 =\hbar^{2g+n-2}$,
which is exactly the  $\hbar$ grading in 
\eqref{genusn}.  Thus a connected random field cumulant is the quantum mechanical carrier of connected bulk topology: its coefficient $C^{[g]}_{\vec m}$ represents the genus-$g$ contribution associated with the chosen insertions.  
This scaling is necessary for a given cumulant channel to contribute at a given genus. 
  Together with the matching of the coefficient functions $C^{[g]}_{  \vec m}$, it provides the required genus organization of the ensemble-averaged Hamiltonian correlators.

\begin{figure}[!t]
\begin{center}
\begin{tikzpicture}[
  x=1cm,y=1cm,
  endpoint/.style={circle,fill=black,inner sep=1.8pt},
  outline/.style={draw=Sedge,line width=0.82pt,line cap=round,line join=round},
  lab/.style={font=\small},
  glab/.style={font=\small}
]


\begin{scope}[shift={(0,8.75)}]
  \coordinate (L) at (0,2.0); \coordinate (R) at (4.2,2.0);
  \path[shade,top color=Slight!88!white,middle color=Smid!58!Slight,bottom color=SmidDark!72!Smid,outline]
    (L)
    .. controls (0.48,1.15) and (1.18,.66) .. (1.98,.57)
    .. controls (2.93,.46) and (3.67,1.02) .. (R)
    .. controls (3.60,1.47) and (2.92,1.24) .. (2.10,1.20)
    .. controls (1.28,1.24) and (.60,1.47) .. (L) -- cycle;
  \begin{scope}
    \clip (L)
    .. controls (0.48,1.15) and (1.18,.66) .. (1.98,.57)
    .. controls (2.93,.46) and (3.67,1.02) .. (R)
    .. controls (3.60,1.47) and (2.92,1.24) .. (2.10,1.20)
    .. controls (1.28,1.24) and (.60,1.47) .. (L) -- cycle;
%
  \end{scope}
  \node[endpoint] at (L) {}; \node[endpoint] at (R) {};
  \node[lab,above=4pt] at (L) {$1$}; \node[lab,above=4pt] at (R) {$1$};
  \node[glab] at (2.10,.03) {$\vv{m}=(1,1)\quad g=g_{\min}=0$};
\end{scope}

\begin{scope}[shift={(8.2,8.75)}]
  \coordinate (L) at (0,2.0); \coordinate (R) at (4.2,2.0);
  \path[shade,top color=Slight!88!white,middle color=Smid!58!Slight,bottom color=SmidDark!72!Smid,outline]
    (L)
    .. controls (.48,1.13) and (1.02,.58) .. (1.96,.48)
    .. controls (2.96,.36) and (3.68,.98) .. (R)
    .. controls (3.63,1.58) and (2.90,1.38) .. (2.10,1.36)
    .. controls (1.30,1.38) and (.57,1.58) .. (L) -- cycle;
  \begin{scope}
    \clip (L)
    .. controls (.48,1.13) and (1.02,.58) .. (1.96,.48)
    .. controls (2.96,.36) and (3.68,.98) .. (R)
    .. controls (3.63,1.58) and (2.90,1.38) .. (2.10,1.36)
    .. controls (1.30,1.38) and (.57,1.58) .. (L) -- cycle;
%
    \pic[scale=1.2] at (2.10,.84) {puncture3d};
  \end{scope}
  \node[endpoint] at (L) {}; \node[endpoint] at (R) {};
  \node[lab,above=4pt] at (L) {$1$}; \node[lab,above=4pt] at (R) {$1$};
  \node[glab] at (2.10,.02) {$\vv{m}=(1,1)\quad g=1$};
\end{scope}

\begin{scope}[shift={(0,5.2)}]
  \coordinate (L) at (0,2.0); \coordinate (R) at (4.2,2.0);
  \path[shade,top color=Slight!88!white,middle color=Smid!58!Slight,bottom color=SmidDark!72!Smid,outline]
    (L)
    .. controls (.48,1.13) and (1.02,.58) .. (1.96,.48)
    .. controls (2.96,.36) and (3.68,.98) .. (R)
    .. controls (3.63,1.58) and (2.90,1.38) .. (2.10,1.36)
    .. controls (1.30,1.38) and (.57,1.58) .. (L) -- cycle;
  \begin{scope}
    \clip (L)
    .. controls (.48,1.13) and (1.02,.58) .. (1.96,.48)
    .. controls (2.96,.36) and (3.68,.98) .. (R)
    .. controls (3.63,1.58) and (2.90,1.38) .. (2.10,1.36)
    .. controls (1.30,1.38) and (.57,1.58) .. (L) -- cycle;
    \pic[scale=1.2] at (2.10,.84) {puncture3d};
  \end{scope}
  \node[endpoint] at (L) {}; \node[endpoint] at (R) {};
  \node[lab,above=4pt] at (L) {$3$}; \node[lab,above=4pt] at (R) {$1$};
  \node[glab] at (2.10,.02) {$\vv{m}=(3,1)\quad  g=g_{\min}=1$};
\end{scope}

\begin{scope}[shift={(8.2,5.2)}]
  \coordinate (L) at (0,2.0); \coordinate (R) at (4.2,2.0);
  \path[shade,top color=Slight!88!white,middle color=Smid!58!Slight,bottom color=SmidDark!72!Smid,outline]
    (L)
    .. controls (.46,1.09) and (1.06,.53) .. (1.96,.43)
    .. controls (2.94,.32) and (3.70,.93) .. (R)
    .. controls (3.60,1.58) and (2.92,1.39) .. (2.10,1.34)
    .. controls (1.28,1.39) and (.60,1.58) .. (L) -- cycle;
  \begin{scope}
    \clip (L)
    .. controls (.46,1.09) and (1.06,.53) .. (1.96,.43)
    .. controls (2.94,.32) and (3.70,.93) .. (R)
    .. controls (3.60,1.58) and (2.92,1.39) .. (2.10,1.35)
    .. controls (1.28,1.39) and (.60,1.58) .. (L) -- cycle;
    \pic[scale=.84] at (1.69,.76) {puncture3dcompact};
    \pic[scale=.84] at (2.51,.76) {puncture3dcompact};
  \end{scope}
     \node[endpoint] at (L) {}; \node[endpoint] at (R) {};
  \node[lab,above=4pt] at (L) {$3$}; \node[lab,above=4pt] at (R) {$3$};
  \node[glab] at (2.10,-.02) {$\vv{m}=(3,3)\quad  g=g_{\min}=2$};
\end{scope}

\begin{scope}[shift={(0,1.5)}]
  \coordinate (L) at (0,2.0); \coordinate (M) at (2.10,2.0); \coordinate (R) at (4.2,2.0);
  \path[shade,top color=Slight!88!white,middle color=Smid!58!Slight,bottom color=SmidDark!72!Smid,outline]
    (L)
    .. controls (.42,1.09) and (1.06,.55) .. (1.96,.43)
    .. controls (2.94,.32) and (3.73,.93) .. (R)
    .. controls (3.58,1.40) and (2.84,1.25) .. (M)
    .. controls (1.36,1.25) and (.62,1.40) .. (L) -- cycle;
  \begin{scope}
    \clip (L)
    .. controls (.42,1.09) and (1.06,.55) .. (1.96,.43)
    .. controls (2.94,.32) and (3.73,.93) .. (R)
    .. controls (3.58,1.40) and (2.84,1.25) .. (M)
    .. controls (1.36,1.25) and (.62,1.40) .. (L) -- cycle;
  \end{scope}
   \node[endpoint] at (L) {}; \node[endpoint] at (M) {}; \node[endpoint] at (R) {};
  \node[lab,above=4pt] at (L) {$1$}; \node[lab,above=4pt] at (M) {$1$}; \node[lab,above=4pt] at (R) {$1$};
  \node[glab] at (2.10,-.10) {$\vv{m}=(1,1,1)\quad g=g_{\min}=0$};
\end{scope}

\begin{scope}[shift={(8.2,1.5)}]
  \coordinate (L) at (0,2.0); \coordinate (M) at (2.10,2.0); \coordinate (R) at (4.2,2.0);
  \path[shade,top color=Slight!88!white,middle color=Smid!58!Slight,bottom color=SmidDark!72!Smid,outline]
    (L)
    .. controls (.42,1.08) and (1.04,.53) .. (1.96,.41)
    .. controls (2.96,.29) and (3.73,.91) .. (R)
    .. controls (3.86,1.61) and (3.53,1.46) .. (3.18,1.44)
    .. controls (2.78,1.42) and (2.43,1.60) .. (M)
    .. controls (1.77,1.60) and (1.42,1.42) .. (1.02,1.44)
    .. controls (.67,1.46) and (.34,1.61) .. (L) -- cycle;
  \begin{scope}
    \clip (L)
    .. controls (.42,1.08) and (1.04,.53) .. (1.96,.41)
    .. controls (2.96,.29) and (3.73,.91) .. (R)
    .. controls (3.86,1.61) and (3.53,1.46) .. (3.18,1.44)
    .. controls (2.78,1.42) and (2.43,1.60) .. (M)
    .. controls (1.77,1.60) and (1.42,1.42) .. (1.02,1.44)
    .. controls (.67,1.46) and (.34,1.61) .. (L) -- cycle;
    \pic[scale=1.20] at (2.10,.79) {puncture3d};
  \end{scope}
  \node[endpoint] at (L) {}; \node[endpoint] at (M) {}; \node[endpoint] at (R) {};
  \node[lab,above=4pt] at (L) {$1$}; \node[lab,above=4pt] at (M) {$1$}; \node[lab,above=4pt] at (R) {$1$};
  \node[glab] at (2.10,-.10) {$\vv{m}=(1,1,1)\quad  g=1$};
\end{scope}
\end{tikzpicture}
\end{center}
\caption{\small Connected cumulants with holes 
for two or three insertions. 
The marked boundary nodes are shown as black dots carrying the powers $m_i$
 of the master field.
The insertions are connected by a  genus-$g$ surface. 
The displayed value of $g_{\min}$ follows from 
\eqref{gmin}.
}
\label{fig:topological-configurations}
\end{figure}
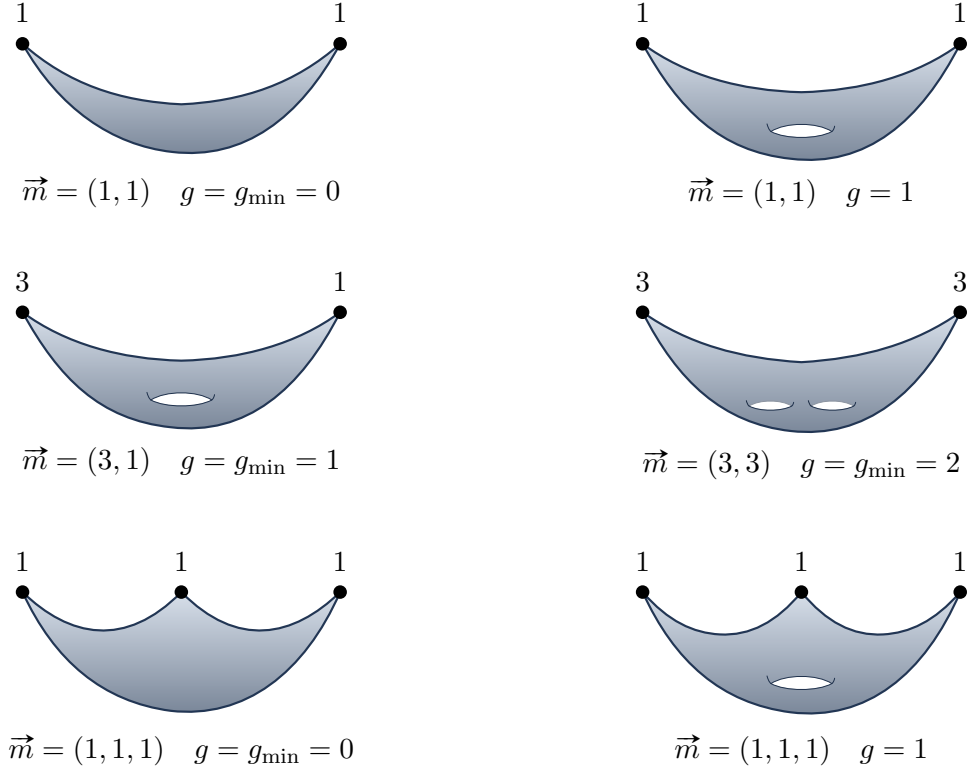
The lower limit $g_{\min}$ follows from a simple but physically meaningful consistency requirement. 
We assume that  the renormalized cumulants are formal power series in $\hbar$ with no  negative powers. Therefore one must impose simultaneously
 $g\geq0$, and $\alpha(g,\vec m)\geq0$.
The first condition is the usual non-negativity of the genus, while the second states that the cumulant itself starts at a nonnegative power of 
$\hbar$. 
Together they give the compact form
\begin{equation}
 g_{\min}(\vec m)
 =\max\left\{0,
\lceil s(\vec m)/2\rceil
-n+1
\right\},
\label{gmin}
\end{equation}
where $\lceil Q\rceil$ denotes the smallest integer greater than or equal to 
$Q$. 
The above is simply the statement that, because higher quantum corrections arrive in steps of $\hbar^2$, an odd deficit of powers requires one additional unit before a regular genus expansion can begin.

Note that \eqref{gmin} gives $g_{\min}$ a transparent interpretation.  Increasing the total composite degree $s(\vec m)$ 
 lowers the required powers of $\hbar$ in the cumulant. If the requested power would be negative at genus zero, the corresponding structure cannot occur on a genus-zero surface and is postponed to the first genus for which the cumulant becomes regular.  Thus $g_{\min}$ is a selection rule: it tells us which topologies are forbidden at low genus before any detailed calculation of the coefficient functions is performed.

We shall use the shorthand notation
\begin{equation}
 \left\langle O_{m_1}(X_1)\cdots O_{m_n}(X_n)\right\rangle^{[g]}
 \equiv
 \hbar^{2g+2n-2-s(\vec m)}
 C^{[g]}_{\vec m}(X_1,\ldots,X_n).
\label{genuscomponent}
\end{equation}
Once the leading allowed genus is known, all higher terms occur at $g_{\min}+1,g_{\min}+2,\ldots$, with successive powers differing by $\hbar^2$.  
At fixed composite field content, the explicit factor $\hbar^{2h}$
 multiplying $A_{2h}$
 shifts the genus grading by $h$.
This closure under ordinary quantum corrections is one of the attractive features of the formulation: once the cumulant scaling~\eqref{cumulantgenus} is imposed, the WK corrections preserve the genus grading at every order.

Here we shall also adopt what we call the mean-field gauge: 
we center the elementary field and choose a renormalization scheme in which the one-point functions of the higher coincident composites vanish\footnote{We continue to denote the centered (shifted) master field by $w$, slightly abusing notation for simplicity.}\negthinspace\negthinspace,
\begin{equation}
 \langle w(X)\rangle=0\,,\qquad
 \langle O_m(X)\rangle=0\quad (m\geq2)\,,
\label{meanfieldgauge}
\end{equation}
and the potential can then be rewritten as 
\be
 W(X)=U(X,\hbar)+\hbar w(X)\,,\qquad
 U(X,\hbar)=W_0(X)+\sum_{g\geq1}\hbar^{2g}u_g(X)\,.
\label{1meanfieldgauge}
\ee
The first condition transfers the elementary one-point response into the deterministic
mean potential $U$ in this mean-field gauge.  The conditions for $m\geq2$ are composite operator subtraction
conventions; they should not be interpreted as additional $\beta$ independent shifts of
$U$.  
This separates the
deterministic spectral data from genuinely connected disorder fluctuations.  
The functions $u_g$ are known from the string equation formulation of the matrix theory \cite{Okuyama:2019xbv,Johnson:2019eik}; below, we shall  describe how they can  be obtained from the present quantum mechanical framework.
In particular, the genus-one case is
\begin{equation}
 u_1(X)=-\frac{1}{12}\,yD^2y\,,
\label{u1mean}
\end{equation}
where $y(u)=dW_0/dX$ and $D=d/du$.

Several examples illustrate the selection rule in Figure \ref{fig:topological-configurations}.  For $\vec m=(1,1)$, $(3,1)$, $(3,3)$, and $(1,1,1)$, one finds respectively
 $g_{\min}=0,1, 2, 0$. We also include the cases $\vec m=(1,1)$ with $g=1$ and $\vec m=(1,1,1)$ also with $g=1$.
Thus the ordinary two-point function of $w$ already contributes at genus zero, whereas sufficiently high composite degree forces the first nonzero term to appear only after some handles are added.  The figure displays these lowest-order possibilities.  Power counting determines when a structure may occur, while the coefficient functions $C^{[g]}_{\vec m}$ 
are constrained by matching the sewn correlators to the corresponding JT amplitudes; individual local representatives may still depend on the composite operator prescription.
This division of labor is useful: topology and $\hbar$ scaling are universal, whereas the detailed local kernels encode the specific JT dynamics.  The full genus-zero matching is developed in the next section.

\section{Genus-zero \texorpdfstring{$n$}{n}-boundary correlations \texorpdfstring{$\mathcal Z_{0,n}$}{Z0n}}
\label{sec:Z0n}

We now prove that the genus-zero multi-boundary amplitudes follow from a  set of local random potential cumulants. Within our prescription~\eqref{Hpert} (see also  Appendix \ref{AppA}), the genus-zero two-point function is fixed by \cite{Bak:2026tvy},
\begin{equation}
 \langle w(X_1)w(X_2)\rangle^{[0]}
 =2W'_0(X_1)\delta(X_1-X_2)\,,
\label{cw1}
\end{equation}
and, in this note,  we take the genus-zero higher primitive  cumulants to be
\begin{equation}
 \left\langle w(X_1)\cdots w(X_r)\right\rangle^{[0]}
 =\hbar^{r-2}2^{r-1}W_0^{(r-1)}(X_1)
 \prod_{a=1}^{r-1}\delta(X_a-X_{a+1})\,,
 \qquad r\geq2\,, 
\label{primitiveZ0}
\end{equation}
where $W^{(k)}_0(X)$ is the $k$-th derivative with respect to $X$.  
The superscript $[0]$ denotes the  genus-zero contribution. 
Here 
coincident powers of $w$ on the same boundary  shall be 
first point-split and composite contractions are then built exclusively from products of the primitive contacts in \eqref{primitiveZ0}. 
The construction below proves that this choice is sufficient to 
reproduce the complete genus-zero boundary amplitudes.

Our goal is to show that the above genus-zero local contacts reproduce, for every $n\geq2$, the known JT result \cite{Saad:2019lba,Turiaci:2024cad,Mertens:2022irh,Okuyama:2020ncd},
\begin{equation}
 \mathcal Z_{0,n}(\beta_1,\ldots,\beta_n)
 =\frac{1}{2\beta_t}
 \left(-
{\partial_X}
\right)^{n-2}
 e^{-\beta_t W_0(X)}\big|_{X=0}\,,
\label{Z0nfinal}
\end{equation}
where $\beta_t=\sum_{i=1}^n\beta_i$.
At first sight this is quite nontrivial.  The expansion of the $n$ traces produces many ways of distributing several $w$ insertions among the boundaries, and the primitive cumulants can join any number of them.  Nevertheless, it turns out that genus-zero power counting forces all connected contractions into a very special class: labelled hypertrees.  An exact count of those hypertrees then removes the separate dependence on the $\beta_i$ and leaves only their sum $\beta_t$.  Finally the resulting contact sum becomes a total derivative  
and the integration collapses to the endpoint $X=0$ as we will see.  At genus zero no positive WK 
term contributes, because every such term carries at least one additional factor $\hbar^2$ and hence starts at positive genus.  Thus only the classical Boltzmann factors need to be expanded below.  We now prove the three statements below in turn.

\subsection{Genus zero as hypertrees}
Expand the classical factor on boundary $i$, $e^{-\beta_i\hbar w(X_i)}$, and 
suppose that its expansion supplies $q_i\geq 1$ insertions of $w$. 
To organize coincident products, we temporarily point-split these insertions 
and treat them as distinct legs attached to the same boundary. The coincidence 
limit is understood after the legs have been assigned to primitive contacts. 
A primitive cumulant joining $r$ such legs will be called an $r$-point 
contact, with $r\geq 2$. 

Consider a connected term containing $m$ primitive contacts. It is useful 
to represent it by a bipartite 
 graph.
The graph has two kinds 
of vertices: the $n$ external boundaries and the $m$ primitive contacts. 
Each $w$ insertion is represented by a line 
 joining its boundary vertex 
to the contact vertex in which it participates. Thus, if 
$ N:=\sum_{i=1}^n q_i $ is the total number of insertions, 
the graph has $n+m$ vertices and $N$ lines.

The genus-zero condition fixes $N$ immediately.  The $n$ traces contribute $\hbar^{-n}$, the $N$ insertions contribute $\hbar^N$, and a contact containing $r$ fields contributes the additional factor $\hbar^{r-2}$ from \eqref{primitiveZ0}.  Since the sum of all contact valences is again $N$, the total power is
\begin{equation}
 -n+N+\sum_{e}(r_e-2)
 =-n+2N-2m,
\end{equation}
where the sum runs over the $m$ primitive contacts and we have used
$N=\sum_e r_e$. 
According to the genus grading in Eq.~\eqref{cumulantgenus}, a connected 
genus-zero $n$-boundary contribution carries the overall weight $\hbar^{n-2}$.
The exponent above must therefore equal $n-2$, which gives
$ N=n+m-1$.
Since the graph is connected, has $n+m$ vertices and $ N=n+m-1$ lines, 
it must be a tree. In particular, multiple lines between the same pair of 
vertices are excluded.

A contact involving $r_e$ distinct boundaries can then be viewed as a 
hyperedge $ e\subset[n]:=\{1,\ldots,n\},\quad r_e:=|e|\geq2, $ namely 
a single contact joining $r_e$ boundary vertices at once. 
The collection of the $m$ contacts thus defines a hypergraph $ H=([n],E(H)).$ 
Such a hypergraph is called a hypertree when the bipartite graph just 
described, with one vertex for each boundary and one for each hyperedge, 
is a tree~\cite{McCammond:2017,McCammond:2004}. Therefore the connected contractions contributing at genus zero 
are precisely labelled hypertrees on the $n$ boundary vertices. 
For a boundary $i$, let $d_i(H)$ denote the number of contacts attached 
to that boundary. Since each point-split insertion gives one incidence line, 
$q_i=d_i(H)$. Also, from $N=n+m-1$, one has 
\begin{equation}
 \sum_{e\in E(H)}(r_e-1)=n-1.
\label{HTcondition}
\end{equation}

The first few cases illustrate the structure. For $n=3$, since the sum in
\eqref{HTcondition} is $2$, one has either a single three-point $(r=3)$ contact 
or two two-point $(r=2)$ contacts.  For $n=4$, there is either a single 
four-point contact, one three-point contact together with 
one two-point contact, or three two-point contacts.
These are not extra assumptions; 
they are simply the partitions of $n-1$ implied by \eqref{HTcondition}.


\subsection{Exact sum over labelled hypertrees}
We now compute the contribution of a fixed labelled hypertree $H$ with
$m$ primitive contacts. For brevity, write $ d_i:=d_i(H)$.
From $N=n+m-1$, one then finds the relation
 $\sum_{i=1}^n(d_i-1)=m-1$.
Boundary $i$ supplies $d_i$ copies of $w$.  For a fixed labelled hypertree the $d_i$ incident contact vertices are distinct, so the factor $1/d_i!$ from the exponential expansion is cancelled by the $d_i!$ assignments of its point-split copies to those contacts.  After including the factor $1/(2\beta_i)$ from each 
rescaled trace, boundary $i$ contributes
$ \frac{1}{2}(-1)^{d_i}\beta_i^{d_i-1}.$
Multiplying over all boundaries therefore gives a numerical factor $2^{-n}$.
Also, each $r_e$-point contact contributes the factor $2^{r_e-1}$
from \eqref{primitiveZ0}. Since $ \sum_e(r_e-1)=n-1$,
the product of all contact factors is $2^{n-1}$.
Combining this with $2^{-n}$ from the boundaries leaves
the universal factor $1/2$.

The delta functions in the primitive contacts identify all boundary
coordinates along the tree, leaving a single integration variable $X$.
The contribution of the fixed labelled hypertree $H$ to
$\mathcal Z_{0,n}$ is then
\begin{equation}
 {\mathcal Z_{0,n}[H]}
 =\frac12(-1)^{n-1+m}
 \left(\prod_{i=1}^n\beta_i^{d_i-1}\right)
 \int_0^\infty dX\,e^{-\beta_t W_0(X)}
 \prod_{e\in E(H)}W_0^{(r_e-1)}(X)\,,
\label{fixedHTcontribution}
\end{equation}
where we use $\sum_i d_i=n+m-1$ for the sign.  
Equation~\eqref{fixedHTcontribution} separates the boundary-degree
dependence from the contact-valence dependence: the former appears only
through $\prod_i\beta_i^{d_i-1}$, while the $X$-dependent factor depends
only on the contact valences $r_e$.


We next sum over all hypertrees on the labelled boundary set
$[n]=\{1,\ldots,n\}$.  To perform the counting, first fix the $m$
contact valences and temporarily label the contact vertices by
$\alpha=1,\ldots,m$, with
$
\sum_{\alpha=1}^m(r_\alpha-1)=n-1.
$
For fixed boundary degrees $d_i\geq1$ and contact valences $r_\alpha$,
the corresponding graph is a labelled bipartite tree.
Its number is given by the bipartite Pr\"ufer formula~\cite{Moon1970,Berge:1989}
\begin{equation}
 N(\boldsymbol d,\boldsymbol r)
 =\frac{(m-1)!}{\prod_{i=1}^n(d_i-1)!}
  \frac{(n-1)!}{\prod_{\alpha=1}^m(r_\alpha-1)!}\,.
\label{bipartiteTreeCount}
\end{equation}
This can be obtained as follows.
A bipartite tree with boundary vertices $1,\ldots,n$ and contact vertices
$1,\ldots,m$ is encoded by two Pr\"ufer words.  The boundary word has
length $m-1$, and the boundary label $i$ appears exactly $d_i-1$ times.
Hence the number of possible boundary words is
$
(m-1)!/\prod_{i=1}^n(d_i-1)!.
$
Similarly, the contact word has length $n-1$, and the contact label
$\alpha$ appears exactly $r_\alpha-1$ times, giving
$
(n-1)!/\prod_{\alpha=1}^m(r_\alpha-1)!.
$
The bipartite Pr\"ufer correspondence is one-to-one, so multiplying the
two factors gives Eq.~\eqref{bipartiteTreeCount}.

Now sum over all allowed boundary degrees. 
When the product of $\beta_i$'s in~\eqref{fixedHTcontribution} is
combined with the first factor in~\eqref{bipartiteTreeCount}, we obtain
\begin{equation}
\sum_{ 
\substack{d_i\geq1\\ \sum_i\negthinspace(\negthinspace d_i\negthinspace- \negthinspace 1 \negthinspace)= m \negthinspace -\negthinspace 1}
}
\negthinspace\negthinspace\negthinspace\negthinspace \negthinspace\frac{\prod_i\beta_i^{d_i-1}}
 {\prod_i(d_i-1)!}
\, =\, \frac{\beta_t^{m-  1}}{(m-1)!}\,.
\label{leftDegreeSum}
\end{equation}
Thus, after summing over the boundary degrees, all dependence on the 
individual boundary lengths reduces to the single combination $\beta_t$.

It remains only to remove the temporary labels on the contact vertices.
Let $k_a$ denote the number of contacts with $r=a+1$. Thus $k_1$ counts 
two-point contacts, $k_2$ three-point contacts, and so on. Since there are $m$
contacts in total,
\begin{equation}
m=\sum_{a\geq1}k_a\,,  \qquad
\sum_{a\geq1}a\,k_a=n-1\,.
\end{equation}
Among the temporarily labelled contacts, permutations of the $k_a$ contacts 
having the same valence describe the same hypertree. Dividing by the 
corresponding factor $\prod_a k_a!$, the total weight of all 
labelled-boundary hypertrees with the fixed contact multiplicities 
$\{k_a\}$ is then given by
\begin{equation} 
\frac{(n-1)!}{\prod_{a\geq1}k_a!(a!)^{k_a}}\, \beta_t^{m-1},
\label{weightedHTcount} 
\end{equation} 
where we combined the second factor in~\eqref{bipartiteTreeCount} 
with~\eqref{leftDegreeSum}.
%

Using \eqref{weightedHTcount} and \eqref{fixedHTcontribution}, and changing from $X$ to $u$, 
the complete genus-zero sum becomes
\begin{equation}
 \mathcal Z_{0,n}
 =\frac12\int_0^\infty\frac{du}{y(u)}\,e^{-\beta_t u}\,
 \mathcal N_n(u,\beta_t)\,,
\label{Z0countedIntegral}
\end{equation}
where
\begin{equation}
\mathcal N_n(u,\beta_t) =
\sum_{\substack{k_1,k_2,\ldots\geq0\\ \sum_{a\geq1} a k_a=n-1}}
(-1)^{n-1+m} \frac{(n-1)!\,\beta_t^{m-1}} {\prod_{a\geq1} k_a!(a!)^{k_a}}
\prod_{a\geq1} \left[(yD)^{a-1}y(u)\right]^{k_a},
\label{countedNumerator}
\end{equation}
%
Thus every labelled hypertree has now been summed exactly.  As a simple
check, for $n=3$ the two possible contact structures, namely two
two-point ($r=2$) contacts and a single three-point ($r=3$) contact,
give $\beta_t y^2$ and $-yDy$, respectively.  Hence
\begin{equation}
\mathcal N_3(u,\beta_t) = \beta_t y^2-yDy\,.
\label{n3HTcheck}
\end{equation}
This provides a simple check of the general counting formula.

%

The remaining sum can be written in closed form.
Since $\left(\frac{d}{dX}\right)^a u=\left(yD\right)^a u$
for $a\geq1$, applying the Fa\`a di Bruno formula to $e^{-\beta_t u(X)}$ gives
\begin{equation}
 (yD)^l e^{-\beta_t u}
 =e^{-\beta_t u}
 \sum_{\substack{k_1,k_2,\ldots\geq0\\
                  \sum_{a\geq1}a k_a\,=\,l}}
 \frac{l!\,(-\beta_t)^m}
 {\prod_{a\geq1}k_a!(a!)^{k_a}}
 \prod_{a\geq1} \left[(yD)^{a-1}y(u)\right]^{k_a}.
\label{FaaZ0}
\end{equation}
For  $l=n-1$, these coefficients agree term by term with the direct hypertree count in 
\eqref{countedNumerator}, after extracting the overall factor $(-1)^{n-1}/\beta_t$. Hence,
\begin{equation}
 \mathcal N_n(u,\beta_t)
 =\frac{(-1)^{n-1}}{\beta_t}\,e^{\beta_t u}(yD)^{n-1}e^{-\beta_t u}\,.
\label{numeratorClosed}
\end{equation}
Note that this can also be written as a simple recursion,
\begin{equation}
\mathcal N_{n+1} 
= -e^{\beta_t u}yD ( e^{-\beta_t u}\mathcal N_n ).
\label{Nrecursion}
\end{equation}
For $n\geq3$, using this relation in~\eqref{Z0countedIntegral} gives
\begin{equation}
\mathcal Z_{0,n} 
= -\frac12\int_0^\infty du\, D\left[
e^{-\beta_t u}\mathcal N_{n-1}(u,\beta_t) \right]\,.
\label{totalderivativeZ0}
\end{equation}
For the JT background and $\beta_i>0$, the contribution from $u=\infty$ 
vanishes.  The bulk integral therefore reduces to the endpoint at $u=0$:
\begin{equation}
\mathcal Z_{0,n} = \frac12\mathcal N_{n-1}(0,\beta_t)\,.
\label{Z0endpoint}
\end{equation}
Using 
\eqref{numeratorClosed}, together with
$yD=d/dX$ and $u=W_0(X)$, this becomes
\begin{equation}
\mathcal Z_{0,n} = \frac{1}{2\beta_t} (-\partial_X)^{n-2}
e^{-\beta_t W_0(X)} \big|_{X=0}\,,
\end{equation}
which is precisely 
\eqref{Z0nfinal}.  The case $n=2$ follows
separately from 
\eqref{cw1} and gives $\mathcal Z_{0,2}=1/(2\beta_t)$.

Within the genus-zero representative used here, the preceding construction
gives a direct route from the primitive cumulants~\eqref{primitiveZ0} to the
multi-boundary amplitudes.  The $\hbar$ power counting fixes
$N=n+m-1$, so every connected genus-zero contraction is represented by a
hypertree.  Summing the corresponding labelled bipartite trees with fixed
contact multiplicities reduces the dependence on the individual boundary
lengths to the single combination $\beta_t$.  The same coefficients appear
in the Fa\`a di Bruno expansion, which converts the remaining bulk
contribution into a total derivative and reduces it to the endpoint value
in~\eqref{Z0endpoint}.

The resulting expression agrees with the standard genus-zero JT amplitudes
\cite{Turiaci:2024cad,Mertens:2022irh,Okuyama:2019xbv}.  This agreement also checks the
normalization and combinatorial factors entering the hypertree sum: changing
the primitive normalization, the $\hbar$ power counting, or the symmetry
factors would alter~\eqref{weightedHTcount} and prevent the reduction to
the endpoint form~\eqref{Z0endpoint}.  Thus, within the minimal diagonal
representative used here, the complete family $\mathcal Z_{0,n}$ is generated
from $W_0$ by the single primitive rule~\eqref{primitiveZ0}, without
introducing additional independent genus-zero contact data.



\section{One-boundary amplitudes at arbitrary genus \texorpdfstring{$\mathcal Z_{g,1}$}{Zg1}}\label{sec:one-boundary}

We next turn to the one-boundary amplitudes.  We begin with genus one, where 
only the terms of relative order $\hbar^2$ are required.  At this order the 
role of the composite operator subtraction can be stated without encountering 
products between higher WK 
coefficients and fluctuating fields.  
We then extend the argument to arbitrary genus, where such products do occur 
and the renormalization has to be applied to the complete local heat-kernel 
expression.

\subsection{Genus one}\label{sec:g1}

For the one-boundary calculation we use the centering and mean-potential 
convention of \eqref{meanfieldgauge}.  To genus one,
\begin{equation}
 U=W_0+\hbar^2u_1+O(\hbar^4),
\end{equation}
while the full potential in a fixed realization is $W=U+\hbar w$.  
Substituting this into \eqref{WKmain} gives
\begin{align}
 Z(\beta) & \cong \frac{1}{\hbar\sqrt{4\pi\beta}}
 \int_0^\infty dX\,e^{-\beta(U+\hbar w) }
 \left[1+\sum_{g\geq1}\hbar^{2g}A_{2g}[U+\hbar w,\beta](X)\right] \notag\\
 &=\frac{1}{\hbar\sqrt{4\pi\beta}}\int_0^\infty dX\,e^{-\beta W_0}\left[
 1-\beta\hbar w
 +\hbar^2\left(-\beta u_1+\frac{\beta^2}{2}w^2 + A_2[W_0,\beta] \right)
 +O(\hbar^3)\right].
\label{Zto_h2}
\end{align}
Note that the fluctuation dependent part of $A_2[U+\hbar w]$ does not appear
at this order. The only fluctuation term is $w^2$, which however gives 
vanishing contribution in the convention \eqref{meanfieldgauge}%
\footnote{One can employ different renormalization conventions that 
redistribute finite local terms between the renormalized composite 
contribution, the mean potential, and the endpoint counterterm.  When the 
corresponding shifts are made consistently, the completed one-boundary 
amplitude is unchanged. Equation~\eqref{meanfieldgauge} is the simplest 
choice among them.\label{footnote2}}.
Using \eqref{appA2}, the coefficient of $\hbar$ in the rescaled trace 
becomes
\begin{equation}
 \mathcal Z_{1,1}(\beta)
 =\int_0^\infty du\,e^{-\beta u}
 \left[-\frac{u_1}{2y}
 -\frac{\beta}{12}Dy+\frac{\beta^2}{24}y\right],
 \label{Z11udensity}
\end{equation}
where we have changed the integration variable from $X$ to $u$ as before.

The last two terms in \eqref{Z11udensity} can be integrated by parts
and we get
\begin{equation}
 \mathcal Z_{1,1}(\beta) = P_1(\beta) + \int_0^\infty du\,e^{-\beta u} R_1(u),
 \label{Z11ibp}
\end{equation}
where
\begin{align}
 P_1(\beta) &= \frac1{24} [ \beta y - Dy ]_{u=0}, \notag \\
 R_1(u) &= -\frac{u_1}{2y}-\frac1{24}D^2y.
\end{align}
Here $P_1$ is a polynomial in $\beta$ whose coefficients are endpoint
values of $y$ and $Dy$, while $R_1(u)$ is a smooth function without explicit
$\beta$ dependence.  
Assuming convergence for every $\beta>0$, the second term in~\eqref{Z11ibp} is the Laplace transform of one fixed smooth bulk function and tends to zero as $\beta\to\infty$. Therefore it cannot  be polynomial unless 
it vanishes identically.
The rescaled JT one-boundary amplitudes $\mathcal Z_{g,1}(\beta)$ with $g\geq1$, however, are
polynomials in $\beta$~\cite{Mirzakhani:2006fta,Saad:2019lba,Mertens:2022irh,Turiaci:2024cad}.
To reproduce the JT amplitudes, it is then necessary to impose the condition
$R_1=0$ which determines $u_1$ uniquely,
\begin{equation}
 u_1=-\frac{1}{12}yD^2y\,,
 \label{u1fromR1}
\end{equation}
which is nothing but \eqref{u1mean}.
The same expression is obtained by expanding the JT matrix-model string 
equation to first nontrivial order~\cite{Saad:2019lba,Okuyama:2019xbv}.  
In fact, this calculation is essentially the same as those appearing in 
the KdV framework of   JT gravity~\cite{Okuyama:2019xbv,Gelfand:1975rn}, once the 
fluctuation term in \eqref{Zto_h2} is set to zero\footnote{A nonzero $c(X)=\langle[w^2]_R\rangle$ adds $\beta c/4y$ to the
density in \eqref{Z11udensity}.  After integration by parts it shifts the
bulk remainder, so that $R_1=0$ determines a correspondingly shifted
$u_1$, and it adds the endpoint term $c(0)/4y(0)$ to $P_1$.  Unless this
term is cancelled by an endpoint counterterm,
the resulting completion differs from standard JT gravity.\label{footnote3}}.

For the JT background in \eqref{LCPot}, the endpoint expansion is
\begin{align}
 X(u)&=\frac{u}{2}+\frac{\pi^2u^2}{4}+\frac{\pi^4u^3}{24}+O(u^4)\,, \notag\\
 y(u)=\frac{du}{dX}
 &=2-2\pi^2u+\frac{3}{2}\pi^4u^2+O(u^3)\,.
 \label{JTendpointjets}
\end{align}
Hence $y(0)=2$, $Dy(0)=-2\pi^2$, and $D^2y(0)=3\pi^4$.  Therefore
\begin{equation}
 {\mathcal Z_{1,1}(\beta)=\frac{\pi^2+\beta}{12}}\,.
 \label{Z11final}
\end{equation}
This agrees with the known JT genus-one one-boundary amplitude.  
Note that, in this calculation, neither $u_1$ nor the WK
correction is a total derivative separately, but their completed sum is. 
The bulk integral then contains no independent information and collapses 
to the endpoint polynomial. 

\subsection{Arbitrary genus $g$}

At higher genus there is an additional issue which does not arise in the $g$=1 case.
The fluctuating field appearing in the Boltzmann factor can multiply the 
fluctuation-dependent parts of the WK coefficients. 
The renormalization must therefore be applied to the complete 
local heat-kernel expression, not separately to its individual factors. 
The object to be renormalized is 
\begin{equation}
 \left[
 e^{-\beta(U+\hbar w)}
 \left(1+\sum_{r\geq1}\hbar^{2r}A_{2r}[U+\hbar w;\beta]\right)
 \right]_R ,
 \label{fullOneBoundaryRenormalization}
\end{equation}
which requires the renormalization of products of composite operators.
Since $[AB]_R\neq\big[[A]_R[B]_R\big]_R$ in general, centering $O_m=[w^m]_R$ alone
does not remove the higher-order one-point terms. 
Finite endpoint counterterms may also be present. Since they do not affect
the smooth bulk remainder discussed below, we suppress them temporarily and 
restore them when the renormalization scheme is fixed.
Then $\mathcal Z_{g,1}(\beta)$ can be written in the form
\begin{equation}
 \mathcal Z_{g,1}(\beta)
 =\int_0^\infty du\,e^{-\beta u}
 \sum_{k=0}^{K_g}\beta^k \K_{g,k}(u),
 \label{Zg1localdensity}
\end{equation}
where $\K_{g,k}$ are local expressions made from $y,Dy,D^2y,\ldots$ and 
from the lower-genus potentials $u_{g'}$ with $g'\leq g$. 
Repeated integration by parts as in \eqref{Z11ibp} gives 
\begin{align}
 \mathcal Z_{g,1}(\beta)
 =&\;P_g(\beta)
 +\int_0^\infty du\,e^{-\beta u}R_g(u),
 \label{Zg1endpointremainder}\\
 R_g(u)=&\;\sum_{k=0}^{K_g}D^k \K_{g,k}(u),
 \label{Rgdefinition}
\end{align}
where $P_g$ is a finite polynomial in $\beta$ whose coefficients consist of
 endpoint values of $y$ and its derivatives, and $R_g$ is a smooth function of
$u$ without $\beta$ dependence.

Suppose $u_1,\ldots,u_{g-1}$ have already been fixed. We expand 
\eqref{fullOneBoundaryRenormalization}
to order $\hbar^{2g}$. At order $\hbar^{2g}$, the new unknown 
$u_g$ can appear only linearly through the Boltzmann factor
\begin{equation}
 e^{-\beta U}
 =e^{-\beta W_0}\left(1+\cdots-\beta\hbar^{2g}u_g+\cdots\right).
\label{Uexp}
\end{equation}
In particular, it enters $\K_{g,k}$ only for $k=0$ in the form
\begin{equation}
 \K_{g,0} (u)=-\frac{u_g(u)}{2y(u)}+\cdots,
 \label{Kg0mean}
\end{equation}
where the universal factor $1/2\beta$ in front of the integral
cancels the linear factor $\beta$ in \eqref{Uexp} and $1/y$ comes from 
the change of  variable from $X$ to $u$.
Then the smooth function $ R_g(u)$ becomes
\begin{equation}
 R_g(u)
 =-\frac{u_g(u)}{2y(u)} +  S_g\big[y;u_1,\ldots,u_{g-1}\big](u)\,,
 \label{Rgtriangular}
\end{equation}
where, in the chosen renormalization scheme, $S_g$ contains the universal 
WK coefficients and lower-genus $u_{g'}$ ($g'<g$).

As argued in Subsection 
\ref{sec:g1}, however, $R_g(u)$ should vanish 
if the amplitude reproduces the JT result which is a polynomial of $\beta$,
\begin{equation}
 R_g(u)=0\qquad (u>0)\,.
 \label{Rgzero}
\end{equation}
Imposing this polynomiality condition determines the new mean correction
$u_g$ algebraically,
\begin{equation}
 u_g(u)=2y(u)S_g\big[y;u_1,\ldots,u_{g-1}\big](u)\,.
 \label{ugrecursiveD}
\end{equation}
This is the direct quantum mechanical version of the recursive determination 
familiar from the string equation.  
The important point is that, once the bulk subtraction prescription is fixed, the recursion follows from one-boundary polynomiality, or equivalently from the absence of the smooth remainder $R_g$. Polynomiality alone does not fix that prescription or the remaining endpoint freedom.
Then the amplitude is simply given
by a polynomial of $\beta$ determined by the endpoint data,
\begin{equation}
 \mathcal Z_{g,1}(\beta)=P_g(\beta)\,.
 \label{Zg1endpointonly}
\end{equation}


The remaining issue is whether the finite local counterterms required to 
renormalize the coincident products in the complete heat-kernel operator 
\eqref{fullOneBoundaryRenormalization}, together with the associated 
mean potential and endpoint redefinitions, can be chosen consistently so that 
$P_g(\beta)$ is the standard JT result. At the level of the formal one-boundary perturbative expansion, we adopt a renormalization prescription that matches 
the renormalized ensemble average of the complete local heat kernel expression to the deterministic JT/KdV expansion~\cite{Okuyama:2019xbv,Gelfand:1975rn}.  To implement this prescription recursively, suppose that the finite counterterms have been fixed through genus $g-1$,   
so that the renormalized one-boundary expansion agrees with the deterministic JT/KdV result at all lower genera. 
At genus $g$, once the finite bulk subtraction terms have been specified within this prescription, the new mean correction $u_g$ enters $R_g$ linearly with the nonzero coefficient shown in \eqref{Rgtriangular}, so the condition $R_g=0$ determines $u_g$. 
Any remaining finite endpoint freedom is then fixed by the matching condition $P_g(\beta)=\mathcal{Z}^{\mathrm{JT}}_{g,1}(\beta)$, yielding the standard JT one-boundary amplitude. Thus, within this matching prescription, a JT compatible 
renormalization of the one-boundary sector can be specified order by order%
\footnote{Within this prescription, and with the bulk subtractions
specified above, no additional independent endpoint counterterm is
actually needed through genus four: for $1\leq g\leq4$ we have checked
that, with $u_g$ determined by $R_g=0$ without using
$\mathcal Z^{\rm JT}_{g,1}$ as input, the endpoint polynomial generated
by integration by parts alone agrees with $\mathcal Z^{\rm JT}_{g,1}$.}.
 We use a finite subtraction convention in which all finite local contributions are 
 incorporated into the matched heat-kernel expression, the mean potential, and the endpoint counterterms.
As explained in footnote~\ref{footnote2}, alternative finite subtraction conventions give the same completed one-boundary amplitude when accompanied by the corresponding redefinitions of the mean potential and endpoint counterterms.

\section{Conclusions 
}\label{sec:conclusions}
 We have formulated  perturbative JT gravity as a stochastic  ensemble of ordinary one-dimensional Hamiltonians with a confining random potential.  The leading profile $W_0$ is chosen to reproduce the disk density of states, while the connected cumulants of a single master field encode the multi-boundary and higher-genus information. 
Within the projected bulk trace prescription of Section \ref{sec:topological-expansion}, we derived the cumulant power counting required by the JT genus grading,
reconstructed the complete genus-zero amplitude $\mathcal Z_{0,n}$, and showed that one-boundary polynomiality determines the mean-potential  corrections  $u_g$ recursively in the chosen renormalization scheme. 
Taken together, these results establish a nontrivial quantum mechanical realization of these two perturbative JT sectors within the projected bulk-trace and renormalization prescriptions adopted here.

The formulation also supplies something that the averaged amplitudes $\mathcal Z_{g,n}$ alone do not make
manifest: for every fixed realization of $W$, it defines a concrete finite-$S_0$
Hamiltonian dynamics with a discrete spectrum and unitary real time evolution.  This is
the setting in which the spectral wall, the late time saturation of the wormhole length,
and the dynamical avoidance of the would-be black hole singularity can be studied in a
quenched description.  The construction may therefore be viewed as a candidate finite entropy dynamical completion of the perturbative JT sectors, rather than merely as a reorganization of JT gravity amplitudes. 

Much remains to be done.  One of the central technical problems is to construct the whole
multi-boundary amplitude $\mathcal Z_{g,n}$ hierarchy with $g>0$ and $n>1$~\cite{Eynard:2004mh,Mirzakhani:2006fta, Eynard:2007kz,Eynard:2007fi}, and to formulate a systematic, regulator controlled calculus of
renormalized composite cumulants whose fully integrated observables  are  independent of the choice of local representative.  
Equally important is the microscopic question of how the stochastic continuum operators appearing in our multi-boundary Hamiltonian formulation are related to finite-$\N$ discrete data, where $\N$ denotes the size of the $\N\times \N$ matrix $M$ in the matrix model. 
A natural intermediate description is provided by finite-lattice quantum mechanics based on random Jacobi matrices~\cite{Dumitriu:2002ntg,EdelmanSutton}.
 It would be important to establish more precisely how finite-$\N$ normalized trace functionals and such discrete local data give rise, in the continuum limit, to the random potential representatives employed in our formulation.
In this regard, the Airy limit of the JT model provides a particularly clean testing ground  for studying the relation between a finite-dimensional random matrix model and continuum stochastic operators~\cite{RamirezRiderVirag, GorinShkolnikov}.  

There could also be several dynamical extensions. In particular, it would be interesting to investigate the role of the confining potential in black hole evaporation, especially in relation to information recovery from Hawking radiation~\cite{Almheiri:2020cfm, Kim:2020cds}. The typicality of singularity resolution~\cite{Susskind:2015toa}, as well as possible applications to cosmological singularities viewed from the black hole interior, also deserves further study. Thus, the present work establishes only a starting point; completing the general amplitude hierarchy and developing its full gravitational interpretation remain important directions for future research.

\section*{Acknowledgement}
DB was supported in part by Basic Science Research Program through NRF funded by
the Ministry of Education (2018R1A6A1A06024977). 
S.-H.Y. was supported in part by NRF grant funded by the Korea
government(MSIT) RS-2025-00553127.

\appendix
{\center \section*{Appendix}}

\section{Perturbative expansion of the thermal trace}\label{app:WK}\label{AppA}

This appendix reviews the WK 
formula used in the main text. 
Throughout this appendix,
\begin{equation}
 Z(\beta)=\mathrm{tr}\,e^{-\beta H}
 =\int dX\,\langle X|e^{-\beta H}|X\rangle
 \label{appZcoordinate}
\end{equation}
denotes the unrescaled thermal trace while we use also  
the rescaled trace $\mathcal Z(\beta)=\sqrt{\pi/\beta}\, Z(\beta)$ in the main text.  

We begin with a smooth 
potential in the general one-dimensional
Hamiltonian 
\begin{equation}
 H=p_X^2+V(X)\,,\qquad p_X=-i\hbar\partial_X\,.
 \label{appGeneralHamiltonian}
\end{equation}
For an operator $\widehat O$, define its Wigner transform by
\begin{equation}
 O_W(X,p)
 :=\int_{-\infty}^{\infty}dy\,
 e^{-ipy/\hbar}
 \left\langle X\negthinspace+\negthinspace\frac{y}{2}\left|\widehat O\right|X\negthinspace -\negthinspace\frac{y}{2}\right\rangle \,.
 \label{appWignerTransform}
\end{equation}
Here $p$ is the phase space momentum variable conjugate to $X$.  The trace property of
the Wigner transform gives
\begin{equation}
 Z(\beta)=\frac{1}{2\pi\hbar}\int dX\,dp\,
 \bigl(e^{-\beta H}\bigr)_W(X,p)\,.
 \label{appPhaseSpaceTrace}
\end{equation}

It is useful to introduce the momentum-integrated kernel
\begin{equation}
 K(\hbar,X):=\int dp\,dy\,
 e^{-ipy/\hbar}
 \left\langle X\negthinspace+\negthinspace\frac{y}{2}\left|e^{-\beta H}\right|X\negthinspace-\negthinspace\frac{y}{2}\right\rangle \,.
 \label{appWignerKernel}
\end{equation}
For fixed $V$ independent of $\hbar$, $H$ is invariant under $\hbar\to-\hbar$.  Changing the integration variable $p\to-p$ therefore gives $K(\hbar,X)=K(-\hbar,X)$.
Using this symmetry and the symmetrized heat equation in the Weyl representation, we can show that, for fixed $V$, the formal bulk expansion of the momentum-integrated thermal trace contains only even powers of $\hbar$ beyond the overall factor $\hbar^{-1}$.
Indeed, the partition function may be written in the standard  WK
form~\cite{Wigner:1932eb,Hillery:1983ms,Brack:1997}
\begin{equation}
Z
= \int \frac{dX dp}{2\pi\hbar}\,
e^{-\beta \big(p^2 + V(X)\big)}
\left[
1 + \sum_{r=1}^{\infty} \hbar^{2r} A_{2r}(X,\beta)
\right]\,,
\label{WKgeneral}
\end{equation}
where the coefficient functions $A_{2r}(X,\beta)$ are local expressions constructed from $V(X)$ and its derivatives and, in the
form used here, are independent of the momentum $p$. 
At order $\hbar^2$, the coefficient is given by the well-known expression
\be
A_2(X,\beta)=\frac{1}{12}\left(
\beta^3(V')^2-2 \beta^2 V''
\right)\,. 
\label{appA2}
\ee
Performing the Gaussian integral over $p$, one 
obtains
\be
Z=\frac{1}{\hbar \sqrt{4\pi \beta}}\int dX e^{-\beta V(X)}\left[
1+\frac{1}{12}\left(
 \beta^3(V')^2-2\beta^2 V''
\right) \hbar^2 +O(\hbar^4)
 \right]\,.
\ee
Higher coefficients may be generated
systematically and some explicit forms 
are given in Refs.~\cite{Brack:1997,Jizba:2014rvg}.

For the JT quantum mechanics, $V$ is the potential in \eqref{QMHamiltonian},
\begin{equation}
 V(X)=e^{-X/\hbar}+W(X)\,.
 \label{appJTPotential}
\end{equation}
For every fixed $X>0$, the exponential term is smaller than any power of $\hbar$ as
$\hbar\to0$, while it becomes an infinitely steep barrier for $X<0$.  The expansion is not
uniform in the boundary layer $X=O(\hbar)$, so the  
WK series alone
does not capture  the wall endpoint contributions. 
 In this paper, we define the perturbative genus sector by the projected bulk-trace prescription of  \eqref{Hpert}: the bulk WK density constructed from $W$ is integrated over $X\geq 0$, with $X=0$ retained as the integration endpoint. This prescription does not define an independent self-adjoint Hamiltonian on the half-line with Dirichlet or Neumann boundary conditions. Separate boundary-layer corrections from the exponential wall and boundary heat-kernel coefficients associated with such boundary conditions are not included in this prescription. Endpoint terms generated by integration by parts of the bulk WK density are retained and should be distinguished from these separate wall and boundary-condition corrections.

With this prescription, the projected bulk trace has the formal all-order expansion
\begin{equation}
 Z(\beta)\cong\frac{1}{\hbar\sqrt{4\pi\beta}}
 \int_0^\infty dX\,e^{-\beta W(X)}
 \left[1+\sum_{r=1}^{\infty}\hbar^{2r}A_{2r}[W,\beta](X)\right],
 \label{appWKhalfLine}
\end{equation}
where $\cong$ denotes equality as a formal series within the projected bulk-trace prescription, rather than equality to the full thermal trace up to nonperturbative terms.
In the mean-field convention of \eqref{meanfieldgauge}-\eqref{1meanfieldgauge},  $W=U+\hbar w$. The complete local heat-kernel expression, including products between the Boltzmann factor and the WK coefficients, is renormalized according to 
\eqref{fullOneBoundaryRenormalization} before ensemble averaging.

\end{document}